\documentclass{article}

\usepackage{url}
\usepackage{graphicx}
\usepackage{amsmath}
\usepackage{soul,color}
\begin{document}
%\onecolumn
%\firstpage{1}

\title{Compaction of granular material inside confined geometries}
\maketitle
%\title[]{On cleaning small tubes}
\author{Benjy Marks, Bj{\o}rnar Sandnes, Guillaume Dumazer, Jon Alm Eriksen and Knut J{\o}rgen M{\aa}l{\o}y}
%\address{}
%\correspondance{}
%\extraAuth{}
%\topic{Flow and Transformations in Porous Media}

\begin{abstract}

%\section{}
In both nature and engineering, loosely packed granular materials are often compacted inside confined geometries. Here, we explore such behaviour in a quasi-two dimensional geometry, where parallel rigid walls provide the confinement. We use the discrete element method to investigate the stress distribution developed within the granular packing as a result of compaction due to the displacement of a rigid piston. We observe that the stress within the packing increases exponentially with the length of accumulated grains, and show an extension to current analytic models which fits the measured stress. The micromechanical behaviour is studied for a range of system parameters, and the limitations of existing analytic models are described. In particular, we show the smallest sized systems which can be treated using existing models. Additionally, the effects of increasing piston rate, and variations of the initial packing fraction, are described.

%\tiny
% \keyFont{ \section{Keywords:} Granular Material, Janssen stress, Boundary effects, Confinement, Deformable media, Hele-Shaw cell, Discrete element method, Micromechanics. } %All article types: you may provide up to 8 keywords; at least 5 are mandatory.
\end{abstract}

\section{Introduction}

When granular materials are placed in confined geometries, we often observe a significant portion of the stress being redirected towards the confining boundaries. This phenomenon has been systematically studied for many systems \cite{brauer2006granular,landry2003confined,sperl2006experiments}, most notably in silos, beginning with \cite{janssen1895versuche}. Force redirection has been attributed to the granular nature of the material, and has in many cases been shown to be well represented by a constant coefficient, known as the Janssen coefficient $K$, defined in one spatial dimension as

\begin{equation}
    K=\sigma_{r}/\sigma_{n}, \nonumber
\end{equation}

\noindent where $\sigma_r$ is the redirected stress due to some applied normal stress $\sigma_n$. Here we investigate the development of stresses within a granular packing, confined between two horizontal plates, subjected to a rigid piston impacting it from one side. As the piston moves, granular material is compacted near the piston, and with increasing displacement of the piston, the size of the packing increases. Such an accumulation process is known to occur in the petroleum industry, where sand is liberated from the host rock during extraction, altering the underground morphology of cracks \cite{tronvoll1994experimental,veeken1991sand}. This may also be relevant for understanding proppant flowback in propped fractures \cite{milton1992factors}. Additionally, this geometry is representative of a number of recent experimental studies in Hele-Shaw cells \cite{eriksen2015bubbles,knudsen2008granular,sandnes2011patterns,sandnes2007labyrinth} where the validity of Janssen stress redirection has not been ascertained.

There are a number of interesting patterns which form when a granular material is displaced by a flexible interface in such a geometry \cite{knudsen2008granular,sandnes2007labyrinth}. The nature of the patterns have been shown to depend on many factors, primarily the initial packing fraction and the rate of displacement \cite{sandnes2011patterns}. For this reason, we here investigate the microstructural and mechanical evolution of such a system under these conditions. To reduce the complexity of the system, we consider only a rigid piston.

We are interested in systems which are highly confined. In common experiments with granular material inside Hele-Shaw cells, there are in general fewer than 20 grain diameters between the two Hele-Shaw plates, typically down to around 5 grain diameters \cite{sandnes2011patterns}. As the confinement increases, i.e.\ as the gap spacing decreases, we expect a transition from three dimensional behaviour towards a behaviour governed by the boundaries, as demonstrated for vertical silos in \cite{bratberg2005validity}. It is then of interest to study the changes that result from increasing confinement. We expect that altering the confinement will affect the force redistribution. A transition may occur for extremely confined systems where such an assumption concerning force redirection may not be valid.

\begin{figure}
    \begin{center}
    \includegraphics[width=0.99\textwidth]{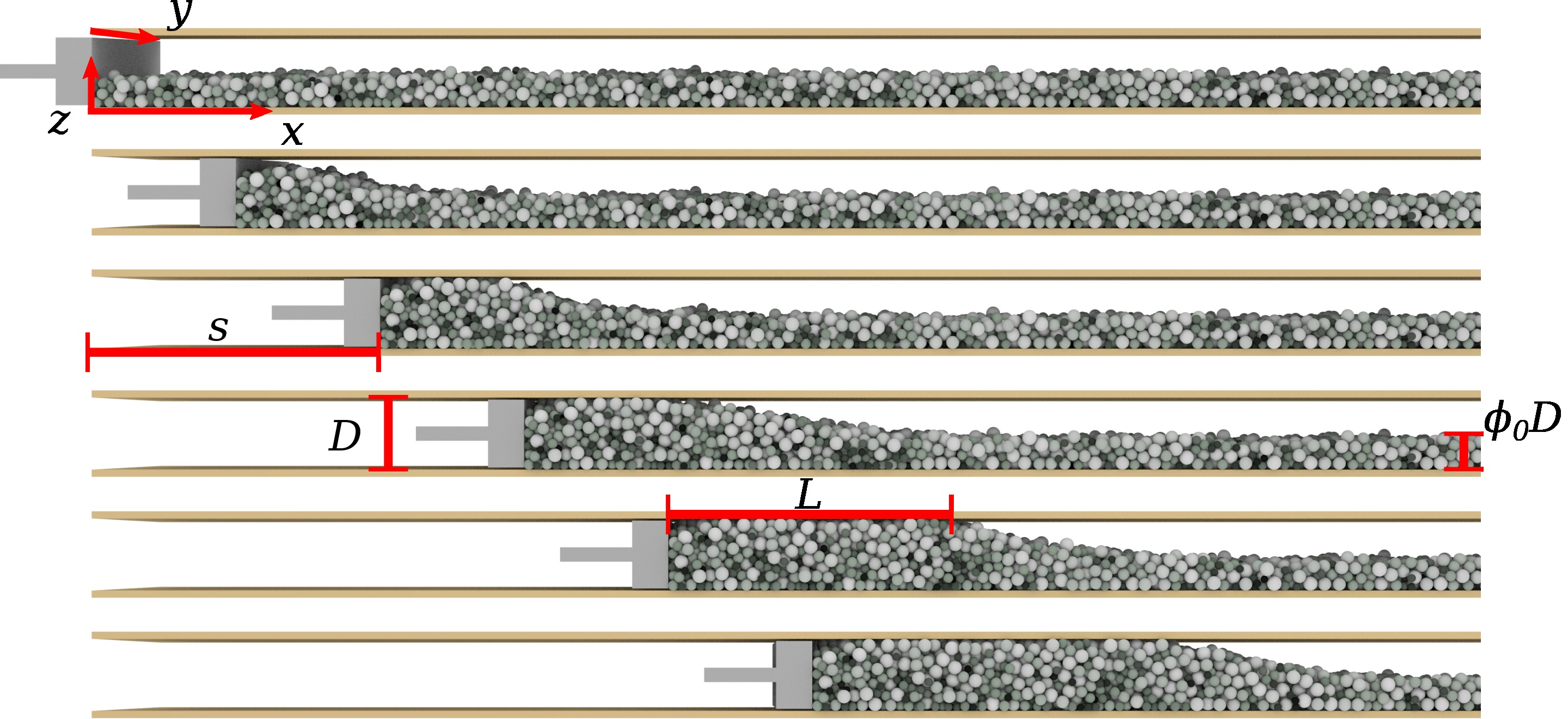}
    \end{center}
    \textbf{\refstepcounter{figure}\label{fig:schematic} Figure \arabic{figure}.}{ Particle positions during a single test, shown at six piston displacements $s$. \emph{Top to bottom}: $s=0,10,20,30,40$ and $50$. Labels refer to the coordinate system $x$, $y$ and $z$, the gap height $D$, the plug length $L$ and the initial packing fraction $\phi_0$. Particles are coloured by size, darker colours representing smaller particles.}
\end{figure}

Existing analytic models for the micromechanics of such a system generally reduce the problem to one spatial dimension ($x$), assuming that variations in both remaining directions ($y$ and $z$) are small, although recently curved interfaces have also been described \cite{eriksen2015bubbles}. For simplicity, we consider a flat interface, and validate the analytic description with a discrete element model.

Firstly, in Section \ref{sec:model} describe the numerical model that has been used to simulate this system. In Section \ref{sec:results}, we establish continuum properties which correspond to the analytic formulation, and show comparisons between the two. In particular, the limitations of current analytic models are identified. Finally, a parameter set is proposed that best fits the analytic theories for a wide range of system variables.

\section{Numerical model}\label{sec:model}

This paper is an investigation into the micromechanics of a system which is highly constrained by external boundaries. For this reason, it is ideal to use a particle based method to model the behaviour, as the total number of grains in the system is small. Towards this end, we use a conventional soft sphere discrete particle approach, implemented in the open source code MercuryDPM (\url{www.mercurydpm.org}) \cite{MercuryDPM1,MercuryDPM2}. The geometry considered here, shown in Figure \ref{fig:schematic}, consists of a rigid piston, oriented in a space $\boldsymbol{r}=\{x,y,z\}$, with normal along the $x$ axis, which pushes particles between two rigid walls, separated by a spacing $D$ and having normals in the $\pm z$ directions, with two periodic boundaries in the remaining perpendicular direction, $y$. The coordinate system moves with the piston, such that it is located at $x=0$ at all times. As the piston moves horizontally at a velocity $u$ towards the grains, its displacement at any time $t$ is then $s=ut$.

We work in a system of non-dimensionalised units with the following properties; length and mass have been non-dimensionalised by the length $d'_m$ and mass $m'_m$ of the largest particle in the system, respectively, where the prime indicates that the quantity has dimension. The particle diameters, $d$, we use are therefore $d\le 1$, with material density defined by $4\pi(1/2)^3\rho_p/3=1$, or $\rho_p=6/\pi$. Time is non-dimensionalised by the time taken for the largest particle to fall from rest its own radius under the action of gravity, so that a unit time is $t=\sqrt{d_{m}/g}$, which requires that $g=1$. Other values are non-dimensionalised by a combination of these three scales, for example stress is non-dimensionalised by $m'_mg'/{d'}^{2}_m$.

Particles are filled into the available space by assigning them to positions on a regular hexagonally close packed lattice, dimensioned such that particles of diameter $1$ would be in contact. In all cases we use particles of $0.5\le d \le 1$ to avoid crystallisation. Variable particle filling is facilitated by changing the number of layers of the grid, such that the initial packing fraction, $\phi_0$ defined in Eq (\ref{eq:phi_0}), is approximately constant throughout the cell. From $t=-10$ to $t=0$, gravity in the $-z$ direction is increased from $g=0$ to $g=1$ to settle the grains in a loose packing. From $t=0$ the piston begins to move at velocity $u$.

\begin{figure}
    \begin{center}
    \includegraphics[width=0.6\textwidth]{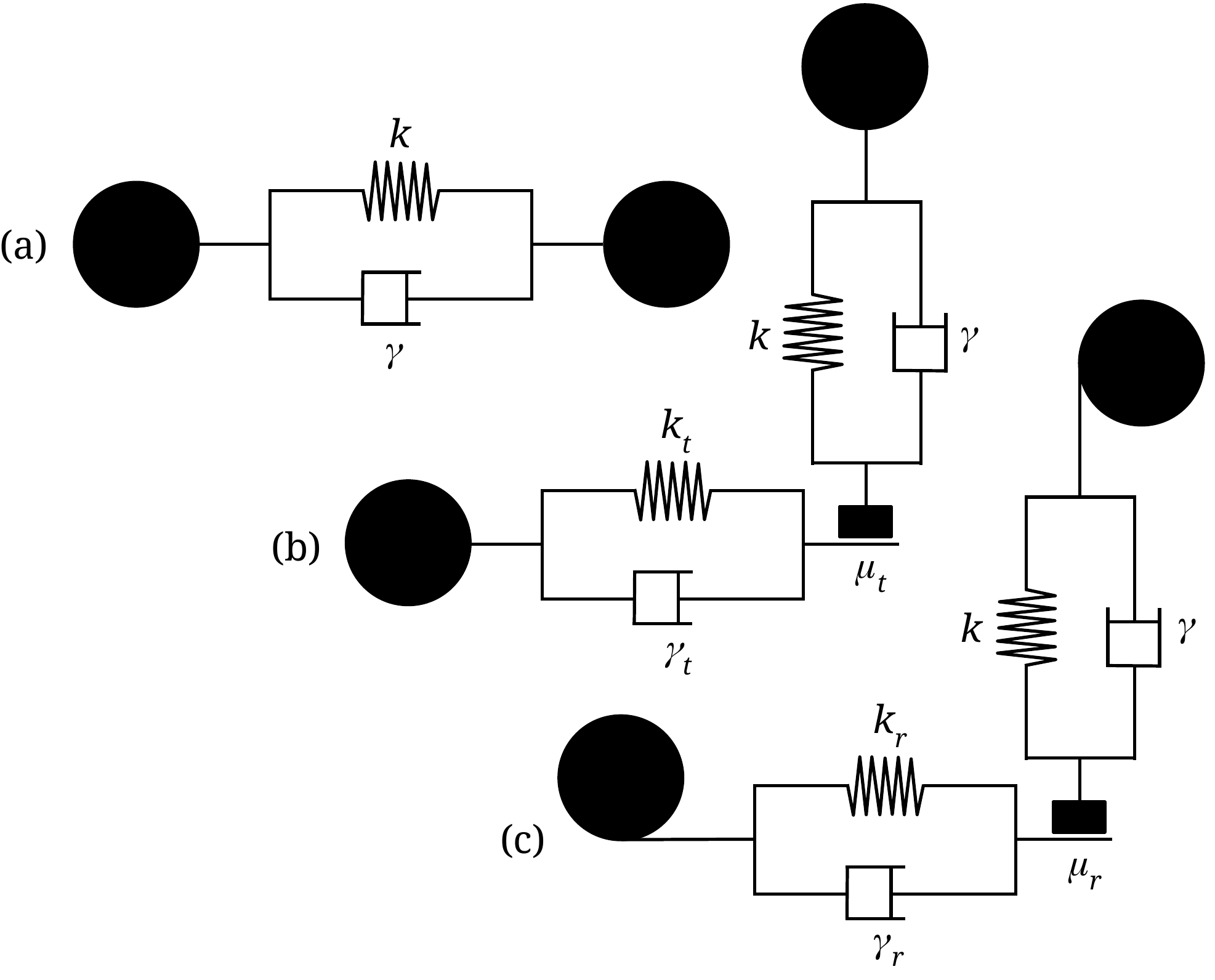}
    \end{center}
    \textbf{\refstepcounter{figure}\label{fig:contact} Figure \arabic{figure}.}{ Contact laws used in the discrete element model. Interactions are characterised by (a) normal, (b) tangential and (c) rolling laws, parameterised by stiffnesses $k$, $k_t$ and $k_r$, viscous dissipation $\gamma$, $\gamma_t$ and $\gamma_r$ and friction coefficients $\mu_t$ and $\mu_r$. Full details are given in \cite{luding2008cohesive}. }
\end{figure}

As shown in Figure \ref{fig:contact}, the particles' material properties are described by normal, tangential and rolling damped spring sliders \cite{luding2008cohesive,MercuryDPM1} with the properties contained in Table \ref{tab:material}. These values have previously been calibrated to mimic micron sized silica beads \cite{fuchs2014rolling}. The walls are implemented such that they are rough; when a particle contacts a wall, the piston, or both, it is prohibited from rotating, i.e.\ $\mu_r = 1$. Otherwise, the interaction properties are the same as between two particles, except that the walls and piston are of infinite mass. We therefore have a well defined macroscopic sliding friction of $\mu=0.4$ that does not depend on the rate of loading.

\begin{table}[!t]
\textbf{\refstepcounter{table}\label{tab:material} Table \arabic{table}.}{ Material properties of the spheres. }

%\processtable{ }
\begin{tabular}{lccc}
Direction & Stiffness, $k$ & Dissipation, $\gamma$ & Coefficient of friction, $\mu$ \\
 & $\left(\frac{\text{kg/s}^2}{m'_m/(g'd'_m)}\right)$ & $\left(\frac{\text{kg/s}}{m'_m/\sqrt{g'd'_m}}\right)$ & \\
Normal & 100000 & 1000 & - \\
Tangential & 80000 & 0 & 0.4 \\
Rolling & 80000 & 0 & 10$^{-3}$ \\
\end{tabular}
\end{table}

In the following Section we will firstly detail the important macroscopic quantities measured from a single simulation. We will then investigate the effect of three controlling parameters on the evolution of the system: the Hele-Shaw spacing $D$, the initial packing fraction $\phi_0$ and the velocity of the piston, $u$. The piston rate, however, is not \emph{a priori} a governing quantity, so we choose to control the piston rate via the inertial number, $I$, which is defined as $I=\dot\gamma d_m/\sqrt{P/\rho_p}$, which is the ratio of inertial to imposed stresses, where $P$ is a typical pressure and $\dot\gamma$ is a typical shear strain rate \cite{midia2004dense}. Taking $\dot\gamma=u/D$, and $P=\rho_p gD$, gives 

\begin{equation}
I = \frac{ud_m}{D^{3/2}g^{1/2}} = uD^{-3/2}. \nonumber
\end{equation}

\section{Results}\label{sec:results}

\begin{figure}
    \begin{center}
    \includegraphics[width=\textwidth]{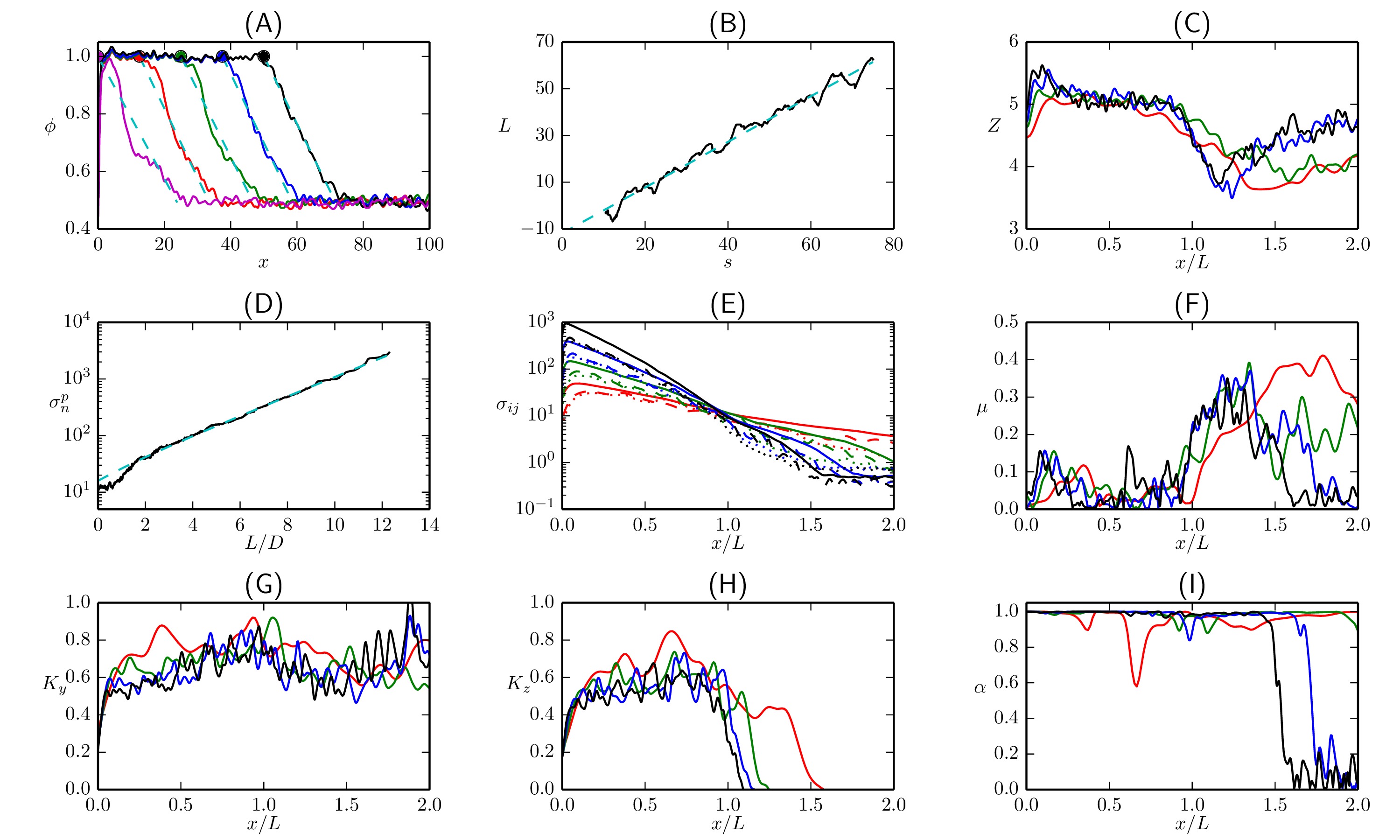}
    \end{center}
    \textbf{\refstepcounter{figure}\label{fig:example} Figure \arabic{figure}.}{ Evolution of coarse grained properties of the system with increasing piston displacement, $s$, for $D=5$, $\phi_0=0.5$ and $I=0.01$.
    \emph{(A)} Five examples of the normalised packing fraction of the particles, $\phi$. The {\color{magenta} magenta}, {\color{red} red}, {\color{green} green}, {\color{blue} blue} and \textbf{black} lines indicate displacements which correspond to $L={\color{magenta} 0},{\color{red} 12.5},{\color{green} 25},{\color{blue} 37.5}$ and $\mathbf{50}$ respectively. The same colour scheme is used for each subsequent plot in this Figure unless stated otherwise. A filled circle indicates the measured value of $L$, and the {\color{cyan} cyan} dashed line indicates the linear best fit measurement of the transition zone. 
    \emph{(B)} Evolution of the measured value of $L$ with increasing displacement $s$ drawn in black. The cyan line indicates the best linear fit to these points.
    \emph{(C)} The coordination number, $Z$, as a function of normalised distance from the piston head, $x/L$. 
    \emph{(D)} Normal stress measured at the piston is shown in \textbf{black}. The {\color{cyan} cyan} dashed line indicates the best fit estimate of Equation (\ref{eq:model}).
    \emph{(E)} Normal stress distributions within the packing. Solid lines indicate $\sigma_{xx}$, dashed $\sigma_{yy}$ and dotted $\sigma_{zz}$.
    \emph{(F)} Apparent friction coefficient measured within the packing, $\mu=|\sigma_{xz}|/\sigma_{zz}$.
    \emph{(G)} Out of plane Janssen coefficient measured within the packing, $K_y=\sigma_{yy}/\sigma_{xx}$.
    \emph{(H)} In plane vertical Janssen coefficient measured within the packing, $K_z=(\sigma_{zz}-\rho_p\nu gD/2)/\sigma_{xx}$.
    \emph{(I)} Absolute value of the $x$ component of the eigenvector of the major principal stress. }
\end{figure}

As depicted in Figure \ref{fig:schematic}, three distinct regions exist inside the system. These are termed the \emph{plug zone}, where particles are densely packed near to the piston ($x\le L$), the \emph{undisturbed zone}, far from the plug, and the \emph{transition zone}, where the plug is accumulating. To define these regions systematically, we must first measure the solid fraction, $\nu =V_s/V_t$, which is a local measure of the ratio of the volume of solids to the total volume. The solid fraction is coarse grained in one spatial dimension, $x$, as

%\[ \nu(\boldsymbol{r},t) = \sum_{i=1}^{N}V_i\mathcal{W}(\boldsymbol{r}-\boldsymbol{r}_i(t)), \]
\begin{equation}\label{eq:nu}
\nu(x,t) = \frac{1}{WD}\sum_{i=1}^{N}V_i\mathcal{W}(x-x_i(t)),
\end{equation}

\noindent where $N$ is the number of grains in the system, $W$ is the width of the system, $V_i$ is the volume of the $i$-th grain, $x_i$ its centre of mass and $\mathcal{W}$ is the coarse graining function, chosen in this case to be a 1D normalised Gaussian function \cite{goldhirsch2010stress,MercuryDPM2}. Such a coarse graining method allows the coarse graining width to be defined, so that either the macro- or micro-structure is visible. We choose to use a coarse graining width equal to the maximum particle diameter, such that small scale variations are minimised, and smooth continuum fields are obtained \cite{weinhart2013coarse}. All other continuum quantities are defined using the same coarse graining technique, presented in \cite{goldhirsch2010stress,MercuryDPM2}. Using the definition (\ref{eq:nu}), we denote the maximum solid fraction, $\nu_m$, as an average over the solid fraction close to the wall at some time when the transition zone is far from the piston as

\begin{equation}
\nu_m = \frac{1}{5D}\int_{5D}^{10D}\nu(x)~dx,
%\nu_m = \frac{1}{5D}\sum_{i=5D}^{10D}\nu_i,
\end{equation}

\noindent where in practice a numerical integration is done over the discrete coarse grained cells, and the limits of $5D$ and $10D$ are chosen arbitrarily.
%\noindent where $A$ and $B$ are the cells containing $x=5D$ and $x=10D$ respectively.
We define the normalised packing fraction, $\phi(x,t)$, as $\phi = \nu/\nu_m$, and note that in the absence of volumetric expansion or dilation of the granular material, $\phi$ is directly proportional to the height of the packing between the Hele-Shaw walls. The undisturbed zone is that which is maintained at the initial packing fraction $\phi_0$, which is defined at time $t=0$ as 

\begin{equation}\label{eq:phi_0}
\phi_0 = \frac{1}{5D}\int_{5D}^{10D}\phi(x,t=0)~dx. \nonumber
\end{equation}

To delineate the plug, transition and undisturbed zones, at each time $t$ we use linear regression to find the best linear fit to the points in the range $\phi=\phi_0 + 0.1$ to $\phi=0.9$, such that $\phi D\approx b - x\tan\theta$, for some value of $b$ and slope angle $\theta$. Five examples of this are shown in Figure \ref{fig:example}A. Using this best fit, we can define the point at which the plug zone meets the transition zone as the intersection of the best fit regression line and $\phi=1$, such that $L=(b-D)/\tan\theta$, as shown in Figure \ref{fig:example}A. The value of $L$ grows monotonically with piston displacement, as shown in Figure \ref{fig:example}B. Conservation of mass implies that on average, if there is no volumetric change in the packing, and no slip between the grains and the walls, this relationship can be expressed as

\begin{equation}\label{eq:Ls}
\frac{L}{s} = \frac{\phi_0}{1-\phi_0},
\end{equation}

\noindent which is shown as the dashed line in Figure \ref{fig:example}B. For all cases reported here, the value of $\theta$ does not appear to vary with increasing plug length $L$. The point at which the undisturbed zone meets the transition zone can then be defined in a similar manner as above, by the intersection of the best fit regression line with $\phi=\phi_0$. The coordination number, $Z$, (average number of contacts per particle), is fairly constant in the plug zone, (Figure \ref{fig:example}C), increasing with compaction, as rearrangement occurs. Additionally, at large values of $L$ the stresses are high enough to cause significant overlap of the particles (up to 1\%). In the transition zone, significant particle rearrangement lowers the coordination number.

Coarse graining techniques in general cause measured fields to converge towards zero near boundaries \cite{MercuryDPM2}. While it is feasible to reconstruct these fields in general near individual boundaries, near the piston we have three distinct boundaries which all interact. To access stresses in this region, it is then preferable to directly measure the forces applied to the rigid boundaries of the system. Towards this end, we denote $\sigma^p_{n}$ as the normal stress measured at the piston. This is shown as a function of piston displacement in Figure \ref{fig:example}D. The stresses measured from coarse graining within the packing are shown in Figure \ref{fig:example}E. In both cases, the stresses grow exponentially both with increasing $L$ and decreasing $x$, as shown by the linearity in a semilog space in Figure \ref{fig:example}D and E.

An analytic expression to describe this stress evolution was first derived in \cite{knudsen2008granular}, assuming that: (a) the stress redirection in the $z$ direction is described by a constant $K_z=\sigma'_{zz}/\sigma_{xx}$, where $\sigma'_{zz} = \sigma_{zz} - D\rho_p\nu_mg/2$ is the component of the vertical stress not due to gravity and (b) friction at the side walls is $\mu=\sigma_{xy}/\sigma_{zz}$. It can be shown using force balance in the $x$ direction that under these conditions, if there is no net acceleration,

\begin{equation}
\frac{d\sigma_{xx}}{dx} = -\frac{2\mu K_z}{D}\sigma_{xx} - \mu\rho_p\nu_mg. \nonumber
\end{equation}

\begin{figure}
    \begin{center}
    \includegraphics[width=0.5\textwidth]{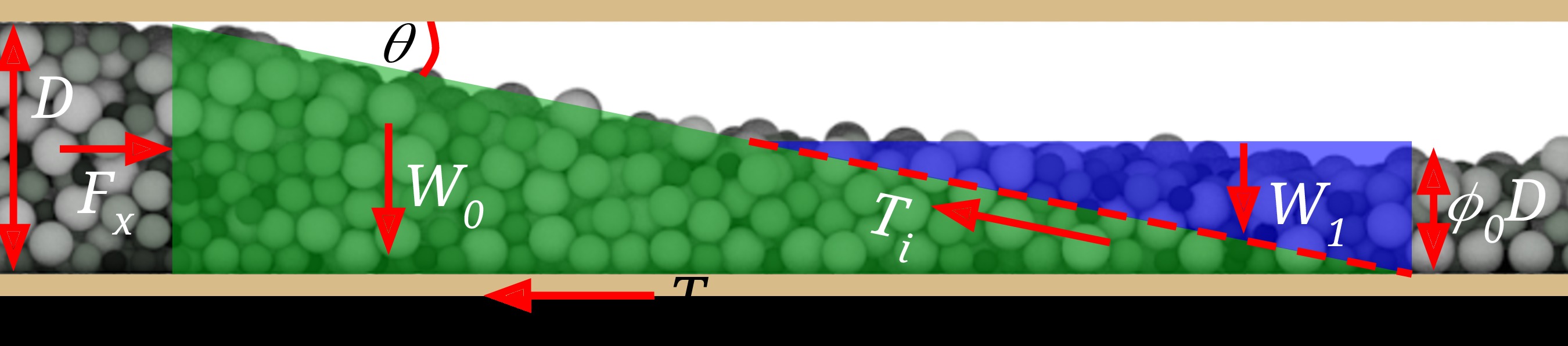}
    \end{center}
    \textbf{\refstepcounter{figure}\label{fig:threshold} Figure \arabic{figure}.}{ Limit equilibrium of the transition zone. For the pile to be displaced by a force $F_x$, two forces must be overcome; $T_b$, the basal traction due to the weight $W_0$ of the green region, and the $x$ component of the internal sliding traction $T_i$ along the assumed failure surface denoted by the dashed line, due to the normal component of the weight above the failure surface, $W_1$, of the blue region.}
\end{figure}

By further assuming that the stress at $x=L$ is a constant, i.e.\ $\sigma_T\equiv\sigma_{xx}(x=L)$,

\begin{equation}\label{eq:almost}
\sigma_{xx} = \left(\sigma_T + \frac{D\rho_p\nu_mg}{2K_z} \right)e^{2\mu K_z(L-x)/D} - \frac{D\rho_p\nu_mg}{2K_z}.
\end{equation}

Previously, the threshold stress $\sigma_T$ has been modelled as either estimated from experimental data \cite{eriksen2015bubbles}, or as a constant by assuming sliding of a wedge of material \cite{knudsen2008granular,sandnes2007labyrinth}. Here, we choose to model the threshold stress $\sigma_T$ as a $\phi_0$ dependent quantity by considering limit equilibrium of a wedge of material being displaced into the undisturbed zone, as shown in Figure \ref{fig:threshold}. We assume a noncohesive Coulomb failure of the material internally, along a failure plane parallel to and meeting the surface of the transition zone. We additionally assume that the internal friction angle is also defined by $\mu$. Limit equilibrium in the $x$ direction can then be expressed using the notation defined in Figure \ref{fig:threshold} as $F_x = T_b + T_i\cos\theta$, where $F_x=D\sigma_T$, $T_b=\mu W_0=\mu D^2\rho_p\nu_mg/(2\tan\theta)$ and $T_i=\mu W_1\cos\theta=\mu D^2\rho_p\nu_mg\phi_0^2\cos\theta/(2\tan\theta)$ per unit length in the $y$ direction. After some rearrangement this implies that

\begin{equation}\label{eq:threshold}
\sigma_T = \frac{\mu D\rho_p\nu_m g}{2\tan\theta}(1 + \phi_0^2\cos^2\theta).
\end{equation}

\noindent This assumption of the failure surface introduces no new parameters into the model, and as will be shown in the following, closely predicts the measured value of $\sigma_T$ for a large range of system parameters. In the limit where $\phi_0\rightarrow0$, this definition reduces to that used in \cite{knudsen2008granular} and \cite{sandnes2007labyrinth}. Including this new definition of the threshold stress, Equation (\ref{eq:threshold}), in Equation (\ref{eq:almost}) gives

\begin{equation}\label{eq:model}
\sigma_{xx} = \frac{D\rho_p\nu_mg}{2K_z}\left(\Big(\frac{\mu K_z}{\tan\theta}(1+\phi^2_0\cos^2\theta) + 1\Big)e^{2\mu K_z(L-x)/D} - 1\right).
\end{equation}

A best fit estimate is used to find $K_z$ from (\ref{eq:model}), and is shown as the dashed line in Figure \ref{fig:example}D at $x=0$, using the measured values of $\nu_m$, $\theta$ and $\phi_0$, which adequately captures the behaviour of the system past $L/D=2$. Before this limit, stress redistribution has not saturated, and $\sigma_{xx}$ is less than the predicted value. We find the same transition value of $L/D\approx2$ for all cases reported here.

The measured value of apparent friction $\mu=\sigma_{xz}/\sigma_{zz}$ inside the packing, shown in Figure \ref{fig:example}F shows that the system is far from failure inside the plug zone, and increasingly unstable in the transition zone. The out of plane Janssen coefficient, $K_y = \sigma_{yy}/\sigma_{xx}$, shown in Figure \ref{fig:example}G, has large variations in $x$, but is not significantly affected by the formation of the plug. The in plane Janssen coefficient (Figure \ref{fig:example}H), $K_z = \sigma'_{zz}/\sigma_{xx}$, however, is strongly influenced by the formation of the plug, and is relatively constant with increasing $L$ inside the plug zone.

\begin{figure}
    \begin{center}
    \includegraphics[width=0.5\textwidth]{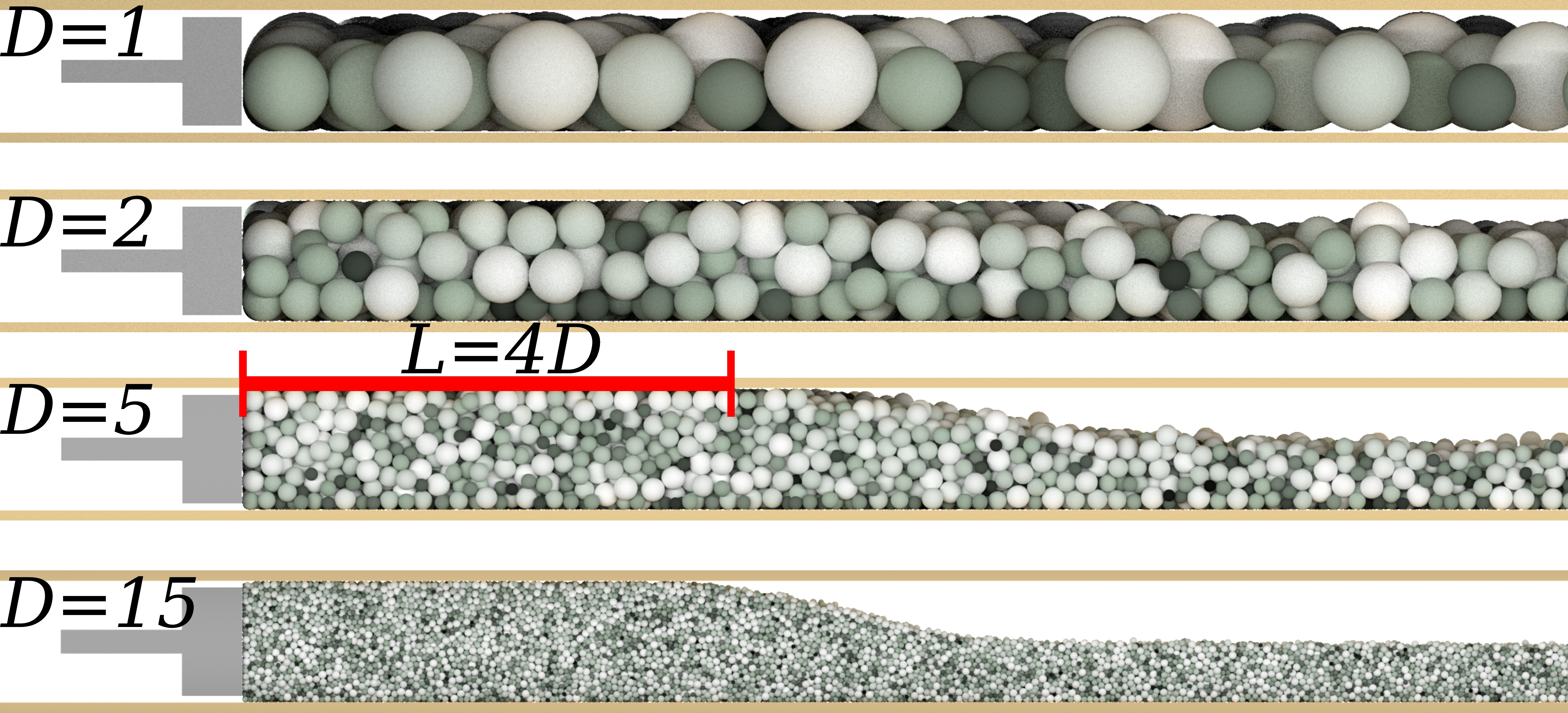}
    \end{center}
    \textbf{\refstepcounter{figure}\label{fig:D_schematic} Figure \arabic{figure}.}{ Particle positions for four simulations with varying gap spacing. \emph{Top to bottom}: $D=1,2,5$ and $15$. For all cases, $\phi_0\approx0.5$, $I=0.01$ and the simulation is displayed at the time corresponding to $L=4D$. Particles are coloured by size, darker colours representing smaller particles.}
\end{figure}

\begin{figure}
    \begin{center}
    \includegraphics[width=\textwidth]{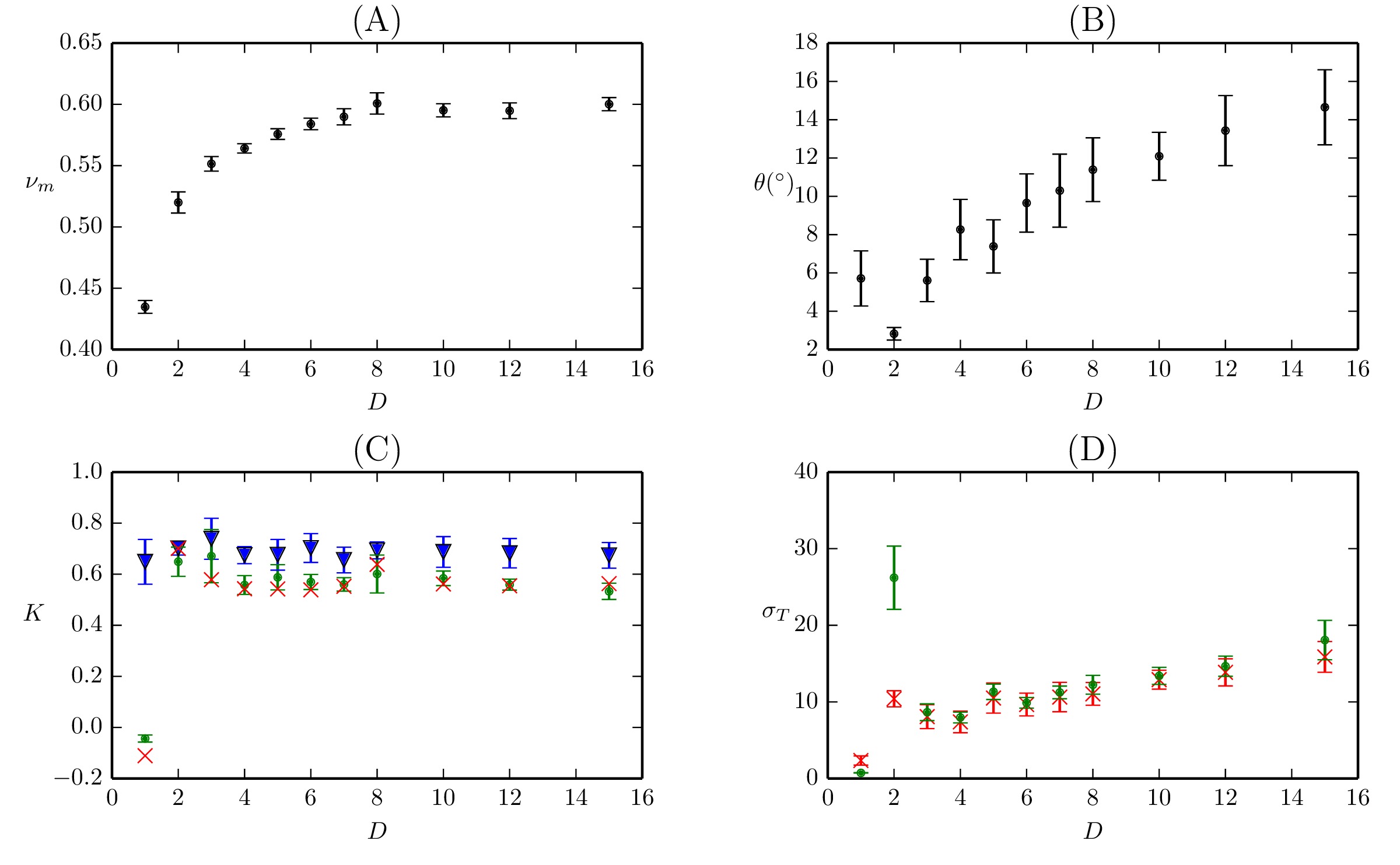}
    \end{center}
    \textbf{\refstepcounter{figure}\label{fig:D} Figure \arabic{figure}.}{ Descriptors of the system with varying gap spacing $D$. Error bars in each plot represent one standard deviation of the measured value.
    \emph{(A)} Average solid fraction within the plug zone, $\nu_m$.
    \emph{(B)} Slope of the pile in the transition zone, $\theta$.
    \emph{(C)} Crosses represent best fit value of $K_z$ from measurement of the stress on the piston head, $\sigma^p_n$ using Eq (\ref{eq:model}). $K_z$ and $K_y$, represented by dots and triangles respectively, are calculated directly from the coarse grained granular packing.
    \emph{(D)} Threshold stress $\sigma_T$. Dots represent the mean value of the coarse grained continuum field $\sigma_{xx}$ at $x=L$, and crosses represent predicted values from Eq (\ref{eq:threshold}) using measured values of $\nu_m$, $\theta$ and $\phi_0$.}
\end{figure}

An underlying assumption of the Janssen stress redistribution is that when averaging over the width of the system (here in the $y$ and $z$ directions), the principal stress directions are parallel to the system geometry \cite{janssen1895versuche}. For this reason we measure $\alpha$, the absolute value of the $x$ component of the eigenvector of the major principal stress, which is shown in Figure \ref{fig:example}I. When $\alpha\approx 1$ the major principal stress points along the $x$-axis, and when $\alpha\approx 0$, the principal stress lies in the $yz$ plane. For $x\le L$ we find that the major principal stress is collinear with the system geometry, and the Janssen stress model fits well. In a traditional silo problem, gravity acts parallel to the applied stress, and averaging across the width of a silo ensures the validity of this assumption. For this case, however, because the direction of gravity has broken the inherent symmetry of the silo problem, its validity is not ensured \cite{nedderman2005statics}. We do, however, find that this assumption holds well for this and all simulations reported here.

\begin{figure}
    \begin{center}
    \includegraphics[width=\textwidth]{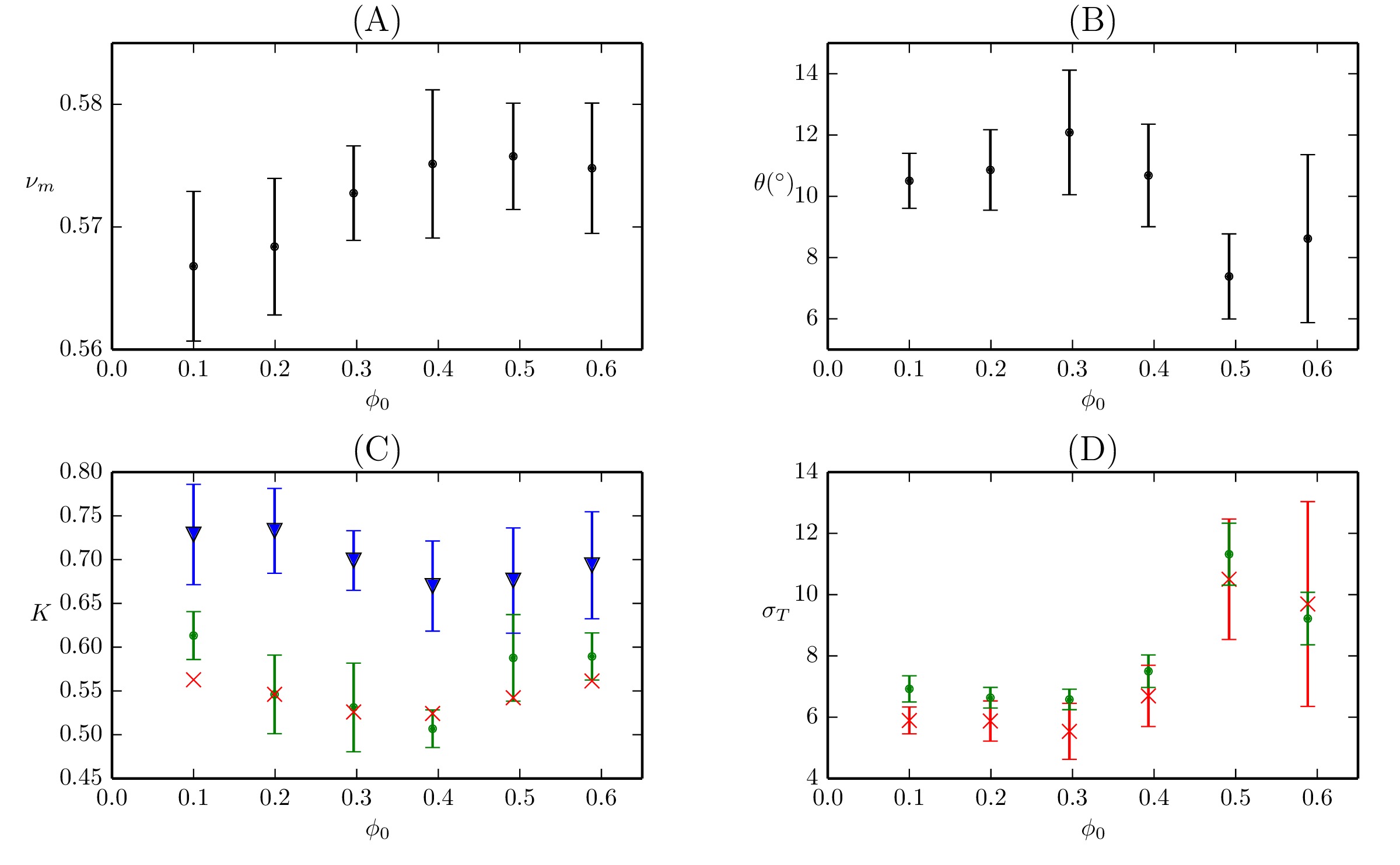}
    \end{center}
    \textbf{\refstepcounter{figure}\label{fig:phi_0} Figure \arabic{figure}.}{ Descriptors of the system with varying initial packing fraction $\phi_0$. Error bars in each plot represent one standard deviation of the measured value.
    \emph{(A)} Average solid fraction within the plug zone, $\nu_m$.
    \emph{(B)} Slope of the pile in the transition zone, $\theta$.
    \emph{(C)} Crosses represent best fit value of $K_z$ from measurement of the stress on the piston head, $\sigma^p_n$ using Eq (\ref{eq:model}). $K_z$ and $K_y$, represented by dots and triangles respectively, are calculated directly from the coarse grained granular packing.
    \emph{(D)} Threshold stress $\sigma_T$. Dots represent the mean value of the coarse grained continuum field $\sigma_{xx}$ at $x=L$, and crosses represent predicted values from Eq (\ref{eq:threshold}) using measured values of $\nu_m$, $\theta$ and $\phi_0$.}
\end{figure}

\subsection{Gap spacing}

As motivated in Equation (\ref{eq:model}), the gap spacing $D$ largely controls the magnitude of the stresses within the system. For this reason, we here vary this spacing systematically from $D=1$ to $D=15$, while maintaining $\phi_0 = 0.5\pm0.05$ and $I=0.01$ (except for the case of $D=1$, where $\phi_0\approx0.66$), as depicted for four values of $D$ in Figure \ref{fig:D_schematic}. To make a reasonable comparison between these simulations, in each case the area of the piston is kept constant, such that its area is $WD = 100$. Select measures of the behaviour of the system are shown in Figure \ref{fig:D}. Slope angles, $\theta$, are averaged over the times corresponding to $2D\le L\le10D$, whilst $K$ and $\sigma_T$ are averaged temporally over values in the range $2D\le L\le 10D$, where at each time we measure spatially in the range $L/4 \le x \le 3L/4$. In Figure \ref{fig:D}A and B, we observe increasing solid fraction in the plug, $\nu_m$, and slope angle, $\theta$, with increasing gap spacing, as the effect of the boundaries on the system decreases. For $D<8$, we note that the effect of the% In Figure \ref{fig:D}D, we observe also that the principal stresses remain co-linear with the geometry for all measured values of $D$.

\begin{figure}
    \begin{center}
    \includegraphics[width=\textwidth]{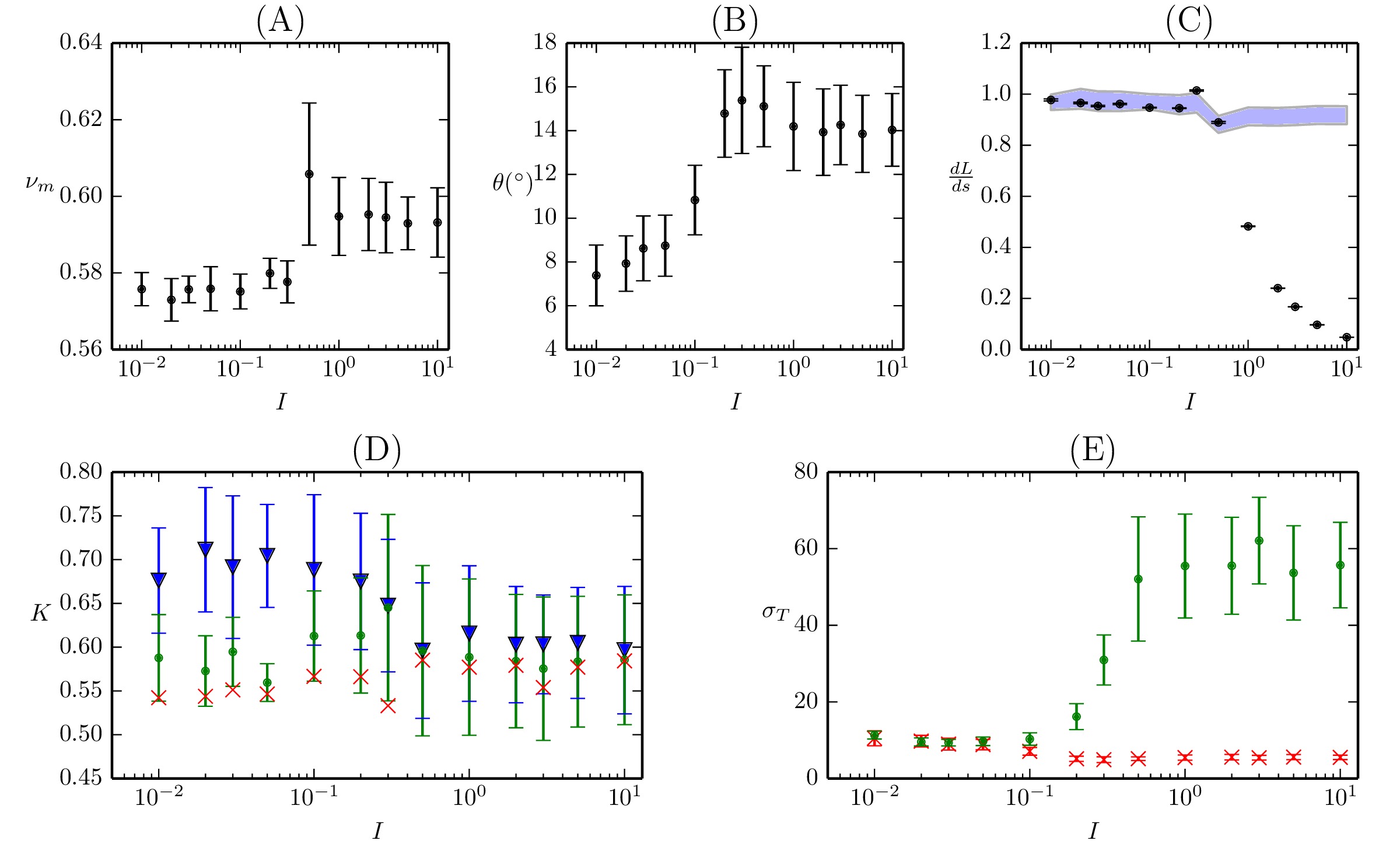}
    \end{center}
    \textbf{\refstepcounter{figure}\label{fig:I} Figure \arabic{figure}.}{ Descriptors of the system with varying inertial number $I$. Error bars in each plot represent one standard deviation of the measured value.
    \emph{(A)} Average solid fraction within the plug zone, $\nu_m$.
    \emph{(B)} Slope of the pile in the transition zone, $\theta$.
    \emph{(C)} Slope of the best fit value of $L(s)$ denoted by black dots against prediction using incompressibilty shown as the shaded region, which denotes  $\pm$ one standard deviation around the mean value for each case.
    \emph{(D)} Crosses represent best fit value of $K_z$ from measurement of the stress on the piston head, $\sigma^p_n$ using Eq (\ref{eq:model}). $K_z$ and $K_y$, represented by dots and triangles respectively, are calculated directly from the coarse grained granular packing.
    \emph{(E)} Threshold stress $\sigma_T$. Dots represent the mean value of the coarse grained continuum field $\sigma_{xx}$ at $x=L$, and crosses represent predicted values from Eq (\ref{eq:threshold}) using measured values of $\nu_m$, $\theta$ and $\phi_0$.}
\end{figure}

For each simulation, the measured normal stress at the piston, $\sigma_n^p$, is fitted with Equation (\ref{eq:model}), and a best fit estimate of $K_z$ is shown as crosses, with the standard deviation of the error of the regression used as error bars, in Figure \ref{fig:D}C. The mean and standard deviation of the measured values of $K_z$ and $K_y$ from the continuum data between $L=2D$ and $L=10D$ are shown as dots and triangles, respectively. For $D\ge2$ we find that both the measured values of $K_z$ and $K_y$ are independent of $D$, and have mean values of $K_y=0.67\pm0.04$ and $K_z=0.58\pm0.05$. Best fit estimates of $K_z$ are also independent of $D$, with mean values of $K_z=0.56\pm0.07$. Figure \ref{fig:D}D shows the mean of $\sigma_T$ also from $L=2$ to $L=10$. Triangles represent the prediction from Equation \ref{eq:threshold} using the measured values of $\nu_m$, $\theta$ and $\phi_0$. In all cases, we find the measured and fitted values of $K_z$ to be in agreement, whereas the values of $\sigma_T$ agree only with $D\ge3$. We note, however, that $\sigma_T$ depends strongly on $\theta$, and we have as yet no means for predicting this quantity. The dependence of $\sigma_T$ on $\theta$ is in contrast to studies on fold and thrust belts \protect\cite{davis1983mechanics}, where there is no confinement vertically above the material.

\subsection{Initial packing fraction}

Existing models \cite{eriksen2015bubbles,knudsen2008granular} for the evolution of the system have neglected any effect of the initial packing fraction $\phi_0$. To test this assumption, we here vary $\phi_0$ from $0.1$ to $0.6$, while maintaining $D=5$ and $I=0.01$. As shown in Figure \ref{fig:phi_0}A and B, the packing fraction inside the plug and the slope of the transition zone are independent of the initial packing fraction. In Figure \ref{fig:phi_0}C, we observe that the measured and best fit values of $K_z$ are in agreement for a wide range of $\phi_0$, and that these values are lower than the measured values of $K_y$. The prediction of threshold stress from Equation (\ref{eq:threshold}), which includes a dependence on $\phi_0$, slightly under-predicts the threshold stress at $\phi_0=0.1$. Nevertheless, both the measured and predicted values of the threshold stress in Figure \ref{fig:phi_0}D are in agreement with observations from \cite{eriksen2015bubbles}, where a non-dimensional threshold stress of $\sigma_T=10.7$ was found to reproduce the observed pattern formation behaviour in the quasi-static limit, at $D=5$, for a range of values of $\phi_0\le0.5$.

\subsection{Inertial number}

Finally, we wish to comment on inertial effects in such a system. Towards this end we systematically vary the inertial number $I$ from $10^{-2}$ to $10$ while maintaining $D=5$ and $\phi_0=0.5\pm0.05$. As shown in Figure \ref{fig:I}, a transition occurs at $I\approx0.1$, where the quasistatic behaviour begins to be dominated by inertial effects, and the system is fluidized. In Figure \ref{fig:I}A, we notice a jump in the maximum solid fraction, as the particles begin to flow and rearrange due to the increased piston velocity. This is accompanied by an increase in the slope angle $\theta$ (Figure \ref{fig:I}B), and a decrease in the accumulation rate (Figure \ref{fig:I}C), as the grains begin to slip relative to the walls. With increasing piston velocity, the anisotropy of the system is lost, as shown in Figure \ref{fig:I}D, and both $K_y$ and $K_z$ tend towards a mean value of $0.56\pm0.02$ for $I\ge1$. As shown in Figure \ref{fig:I}E, the threshold stress, $\sigma_T$, also diverges above $I=0.1$ away from the theoretical prediction. For values of $I\ge0.1$, we have therefore used the measured value of $\sigma_T$ in the best fit estimation of $K_z$ shown in \ref{fig:I}D, rather than the value predicted from (\ref{eq:threshold}), as used in all other cases.

\section{Conclusions}

We have here described a large number of simulations of granular material which have been compacted in a confined geometry. For all cases, we observed that the stress distribution within the packing is well approximated by previous models, once a more rigorous definition of the threshold stress is used. This is true for a wide range of gap spacings, initial filling fractions and piston rates.

In this study we have used a rough boundary condition, where macroscopic friction at the piston and walls is equal to the inter-particle friction. However, in many systems we expect the roughness at the boundaries to be lower than that between particles. It is unclear how this difference will affect either the accumulation of material near the piston head, or the stress distribution within the packing.

Below a gap spacing of 3 particle diameters, the stress distribution is not well represented by this model. We conclude that $D=3$ represents the smallest system size which may reasonably be described by the one dimensional Janssen stress model. In addition, at inertial numbers of $I\ge0.1$, we find that there is significant slip at the boundary, and the threshold stress diverges from the model prediction. In all cases, we cannot as yet predict the slope of the free surface in the transition zone, but we observe that this slope approaches the angle of repose for large systems at low piston rates. Janssen stress coefficients for this system are well represented by $K_z=0.6\pm0.1$ and $K_y=0.7\pm0.1$ for a wide range of system parameters. A model for the threshold stress has been presented using limit equilibrium, and this holds well for systems with $D\ge3$ and $I<0.1$.

The slope angle, $\theta$, has been measured for different system parameters to lie in the range of $2^\circ$-$18^\circ$. \emph{A priori}, we could only assume that this angle must be less than or equal to the angle of repose, which for these grains is approximately $20^\circ$. The wide variability of $\theta$ is as yet unexplained, and is in stark contrast to the case where a top boundary does not exist, for example in fold and thrust belts \cite{davis1983mechanics}. We do note, however, that at large values of $D$ the slope angle approaches the angle of repose.

With regards to the two Janssen parameters, we can clearly distinguish the values of $K_y$ and $K_z$ in Figure \ref{fig:D}C, Figure \ref{fig:phi_0}C and Figure \ref{fig:I}D, for $I<0.1$. The reason for the difference between these two quantities may either be due to anisotropy in the granular packing, or due to the differing boundaries in the $y$ and $z$ directions. As the Janssen parameters are relatively insensitive to the gap spacing $D$, we conclude that this anisotropy is due to the accumulation process, which creates a preferential direction within the packing. This distinction is important when considering models which account for more complex geometries, such as in \cite{eriksen2015bubbles}.

\section*{Disclosure/Conflict-of-Interest Statement}

The authors declare that the research was conducted in the absence of any commercial or financial relationships that could be construed as a potential conflict of interest.

\section*{Author Contributions}
%When determining authorship the following criteria should be observed:
%•	Substantial contributions to the conception or design of the work; or the acquisition, analysis, or interpretation of data for the work; AND
%•	Drafting the work or revising it critically for important intellectual content; AND
%•	Final approval of the version to be published ; AND
%•	Agreement to be accountable for all aspects of the work in ensuring that questions related to the accuracy or integrity of any part of the work are appropriately investigated and resolved.
%Contributors who meet fewer than all 4 of the above criteria for authorship should not be listed as authors, but they should be acknowledged. (http://www.icmje.org/roles_a.html)

BM conducted the simulations, post-processing and authored the paper. BS conceived the idea for the simulations. BS, GD, JAE and KJM assisted with the interpretation and analysis of the simulations, and editing of the manuscript.

\section*{Acknowledgement}

\paragraph{Funding:} The authors would like to acknowledge grants 213462/F20 and 200041/S60 from the NFR.

\bibliographystyle{plain}
\bibliography{bib}

%\section*{Figures} % NEED TO MOVE FIGURES HERE FOR PUBLICATION

\end{document}